\documentclass[11pt]{amsart}
\usepackage{geometry}                
\usepackage{graphicx}
\usepackage{amssymb}
\usepackage{epstopdf}
\usepackage{array}
\usepackage{subfigure}
\usepackage{url}
\usepackage{hyperref}
\usepackage{xspace}
\usepackage{lscape}

\usepackage{color}

\definecolor{gray}{rgb}{0.7,0.7,0.7}

\usepackage{supertabular}
\usepackage{hhline}
\newcommand\arraybslash{\let\\\@arraycr}
\makeatother
\newcommand{\One }{New York area\xspace}

\newcommand{\Two }{Jacksonville area\xspace}
\newcommand{\Three }{Chicago  area\xspace}
\newcommand{\Four }{Dallas  area\xspace}
\newcommand{\Five }{Houston area\xspace}
\newcommand{\Six }{Denver area\xspace}
\newcommand{\Seven }{Albuquerque area\xspace}
\newcommand{\Eight }{Phoenix area\xspace}
\newcommand{\Nine }{Los Angeles area\xspace}
\newcommand{\Ten }{San Jose area\xspace}

\title[Bio-imitation of migration routes]{Bio-imitaiton of Mexican migration routes to the USA  with slime mould on 3D terrains}
\author[Adamatzky]{Andrew Adamatzky}
\address[Adamatzky]{Department of Computer Sciences, University of the West of England, Bristol, United Kingdom}
\email{andrew.adamatzky@uwe.ac.uk}
\author[Martinez]{Genaro J. Mart{\'i}nez}
\address[Mart{\'i}nez]{Escuela Superior de C\'omputo, Instituto Polit\'ecnico Nacional, M\'exico. Department of Computer Sciences, University of the West of England, Bristol, United Kingdom} 
\email{genaro.martinez@uwe.ac.uk}

\markleft{xxxxxxx}

\begin{document}

\maketitle

\vspace{0.5cm}

{\scriptsize
\noindent
Final and edited version of this paper is published in 
Adamatzky A. and Martinez G. J. Bio-Imitation of Mexican Migration Routes to the USA with Slime Mould on 3D Terrains.
J Bionic Engineering  10 (2013) 242--250.
}

\vspace{0.5cm}

\begin{abstract}
Plasmodium of  \emph{Physarum polycephalum} is a large single cell visible by unaided eye. It shows sophisticated behavioural traits in foraging for nutrients and developing an optimal transport network of protoplasmic tubes spanning sources of nutrients. When placed in an environment with distributed sources of nutrients the cell 'computes' an optimal graph spanning the nutrients by growing a network of protoplasmic tubes. \emph{P. polycephalum} imitates 
development of man-made transport networks of a country when configuration of nutrients represents major urban areas. We employ this feature of the slime mould to imitate mexican migration to USA. The Mexican migration to USA is the World's largest migration system.  We bio-physically imitate the migration using slime mould \emph{Physarum polycephalum}.   In laboratory  experiments with 3D Nylon terrains of USA we imitated development of migratory routes from Mexico-USA border to ten urban areas with high concentration of Mexican migrants. From results of laboratory experiments we extracted topologies of migratory routes, and highlighted a role of elevations in shaping the human movement networks. 

\end{abstract}

\section{Introduction}

Natural systems are renown for their abilities to reach optimal solutions of complex dynamical problems. They are 
exploited in the fields of bionics, bioengineering and biocomputing to develop emerging mechanics of information processing and  uncover novel designs of engineering systems. Theoretical part of the bio-inspired information processing field is very well elaborated~\cite{calude_1998,calude_2005}. Experimental part is rather underdeveloped. There is 
just a few of experimental laboratory protypes, e.g. maze-solving micro-fluidic circuits~\cite{Fuerstman2003},
molecular logical gates and circuits~\cite{stojanovic_2002, lederman_2006}, chemical reaction--diffusion processors~\cite{adamatzkyRDC}. Experimental prototypes of novel computing devices are so rare because most novel computing substrates used are somewhat difficult to experiment with.  

When in 2001 Nakagaki with colleagues published a paper on shortest path approximation with live plasmodium of \emph{Physarum polycephalum}~\cite{nakagaki_2001}, and then three years later Tsuda, Aono and Gunji  implemented Boolean logical gates with live slime  mould~\cite{tsuda_2004},  the scientific community got a unique living computing and actuating substrate, which is easy to cultivate and handle, monitor and experiment with.   Slime mould \emph{P.  polycephalum} is an ideal biological substrate for developing biocomputing and bioengineering devices, because the slime mould is  `simpleÕ enough to be studied as spatially extended non-linear media yet robust and  rich behaving to implement a wide range of computational and actuating procedures~\cite{adamatzky_physarummachines, miranda_2011, adamatzky_2008}.

Plasmodium is a vegetative stage of acellular slime mould \emph{P. polycephalum}, a single cell with many nuclei, which feeds on microscopic particles~\cite{stephenson_2000}. When foraging for its food the plasmodium propagates towards sources of food, surrounds them, secretes enzymes and digests the food; it may form a congregation of protoplasm covering the food source. When several sources of nutrients are scattered in the plasmodium's range, the plasmodium forms a network of protoplasmic tubes connecting the masses of protoplasm at the food sources. A structure of the protoplasmic networks is apparently optimal, in a sense that it covers all sources of nutrients and provides a robust and speedy transportation of nutrients and metabolites in the plasmodium's body~\cite{nakagaki_2001}.

Plasmodium's foraging behaviour can be  interpreted as a computation~\cite{adamatzky_physarummachines} as follows.  Data are represented by spatial configurations of attractants and repellents.  Results are represented by structure of protoplasmic  network.  
Plasmodium can solve computational problems with natural parallelism, e.g. related to shortest 
path~\cite{nakagaki_2001} and hierarchies of planar proximity graphs~\cite{adamatzky_ppl_2008}, computation of plane tessellations~\cite{adamatzky_physarummachines,shirakawa, shirakawa_2011}, execution of logical computing schemes~\cite{tsuda_2004, adamatzky_parco}, planar shapes and concave hulls~\cite{adamatzky_parco}, 
and natural implementation of spatial logic and process algebra~\cite{schumann_adamatzky_2009}. 

In previous issues of \emph{Journal of Bionic Engineering} we demonstrate how slime mould \emph{P. polycephalum} could be used to intelligently manipulate objects on a water surface~\cite{adamatzky_2008} and generate sensible 
sound patterns~\cite{miranda_2011}.  The slime mould's behaviour also inspired a range of software implementations of novel approaches towards design of communication and transport networks~\cite{jones_2011,becker_2011}.

When inoculated in an environment with scattered sources of nutrients the slime mould spans the nutrients with a network of protoplasmic tubes resembling human-made transport networks.  Previously, we have conducted a large-scale experimental study to uncover analogies between biological and human-made transport networks and to project behavioural traits of biological networks onto development of vehicular transport networks, see collection of results in \cite{adamatzky_bioevaluation}. Using living slime mould we imitated major road networks in Africa, Australia, Belgium, Brazil, Canada, China, Germany, Iberia, Italy, Malaysia, Mexico, The Netherlands, UK, and USA. We found that for all regions studied~\cite{adamatzky_bioevaluation}, networks of protoplasmic tubes grown by plasmodium match, at least partly, the networks of man-made transport routes. The shape of a country and the exact spatial distribution of urban areas, represented by sources of nutrients, may play a key role in determining the exact structure of the plasmodium network.

Being encouraged by results of imitating the transport networks with slime mould we decided to undertake an analog modelling of human migration. The analog modelling devices built of the slime mould imitates large-scale migration of humans via analogical propagation and foraging of living slime mould. We used Mexican migration to the USA as benchmark task the Physarum analog modeller.  
 
A migration from Mexico to the USA  is the largest people-flow system in the World~\cite{massey_2005}. Every third of Mexicans has been in USA at some stage of their life. The scale of migration is unprecedented. Over eighty percent of migrants are undocumented migrants (there are about  6.2 million of undocumented Mexican migrants amongst  estimated 11 millions of all undocumented migrants in the US)~\cite{passel_2006}. There are several reasons for the migration. First reason is that culture of migration is deeply rooted in Mexican society~\cite{wilson_2010}. Second is that even despite one and a half century passed after Treaty of Guadalupe Hidalgo there remains a high degree of `internal' migration to North Mexican territories detached in the result of the Treaty. Three more reasons for the migration are highlighted in~\cite{massey_espinosa_1997}: market consolidation, social capital formation (people related to USA migrated are more likely to migrate), and human capital formation (if some has previous experience of migration he will more likely to migrate again). 

The migrations shapes economical and social structures of the USA and thus investigation of migratory patterns becomes an issue of great importance. A range of theoretical and analytical models are developed. The models provide a predictive analysis of migration~\cite{massey_zenteno_1999,graves_knapp_1984}, statistical estimates~\cite{colussi_2004,rivero_fuentes_2005},  evaluate integral patterns of migration~\cite{hanson,chang_2010}. Such models make a priceless contribution towards deciding on what measures of controlling migration should be taken, e.g. intensity of a border control might affect migration levels~\cite{eichler_2010} and duration of circular migrants staying in the USA~\cite{thom_2007}.  Most models of Mexican migrations deal rather with integral numbers of migrants, sometimes attributed to particulars states, however neither of models published considered spatial geographic routes of migration.

In present we paper we apply our experience on bio-physical imitation routes~\cite{adamatzky_bioevaluation} to imitate Mexican migration. Also, we go a step further experimentally, comparing to our previous works, --- we take into account configuration of elevations and grow slime mould on a 3D terrain of the USA. We discuss functioning of the Physarum analog modeller, which consists of a living slime mould, 3D plastic model of USA and source of nutrients. Usage of 3D terrain in laboratory experiment bring the analogous modelling closer to reality. The slime mould reacts to 3D topography because \emph{P. polycephalum} is gravisensitive and positively geotropic~\cite{block_1986,block_1998}.  Plasmodium shows morphological geopolarity~\cite{block_1989}: the ectoplasmic wall of a slime mould tube to earth.  We expected that being placed on a 3D terrain with a source of nutrients slime mould propagates towards the source of nutrients and navigates around elevations due to positive geotropism and relatively lower humidity of the elevations.

\section{Methods}

\begin{figure}[!tbp]
\centering
\includegraphics[width=0.7\textwidth]{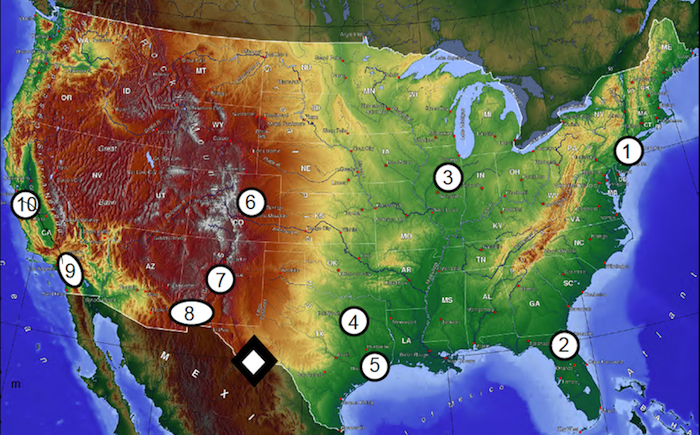}\\
\centerline{(a)}
\includegraphics[width=0.49\textwidth]{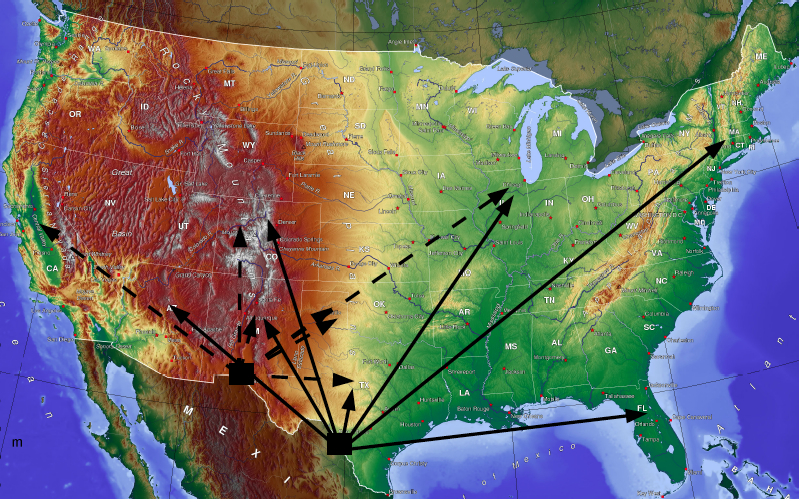}\\
\centerline{(b)}
\includegraphics[width=0.49\textwidth]{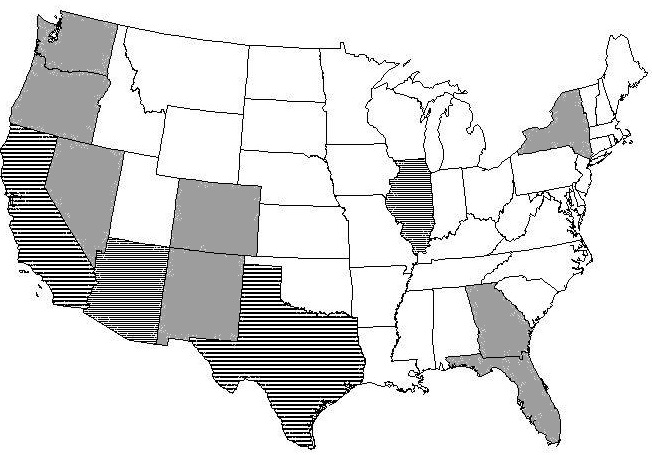}
\centerline{(c)}
\caption{(a)~Configuration of attractor-sites. Site of plasmodium inoculation  is shown by rhomb. (b)~Scheme of major migration routes from Ciudad Ju\'{a}rez, and Nuevo Laredo, generalised from maps~\cite{maps}. (c)~Scheme of state proportion of Mexican born population in the USA, stripes 21\%-37\%, fine stripes 3\%-7\%, grey 1\%-3\%,  modified from \cite{population}.}
\label{methods}
\end{figure}

Terrain of USA used in our experiments is 20~cm wide, 10.8~cm long and 5~cm high. 3D terrain of USA was ordered in \url{http://www.printablegeography.com/}.  The terrain was produced as follows.  The elevation data are downloaded from DIVA-GIS (\url{http://www.diva-gis.org/gdata}), original source is CGIAR (\url{http://srtm.csi.cgiar.org/}). The data are projected with Mercator, and the terrain is  printed using  Selective Laser Sintered PA 2200 with Nylon 12.  To highlight a role of 3D terrain in defining migratory patterns  we conducted experiments on a flat agar plate shaped as USA. Solution of 2\% Aldrich Select Agar was poured in 12~cm square Petri dishes, allowed to cool and then shaped of USA were cut out. We have conducted 14 experiments on flat agar plates and 12 experiments on 3D terrain. It takes slime 3-5 days to colonise all source of nutrients on flat agar, and 5-10 days to colonise the nutrients on 3D terrain.

We have placed oat flakes, to act as sources of nutrients and attractants, in the sites of the 3D terrain  and flat agar plate corresponding to the following areas (Fig.~\ref{methods}a): 

\begin{enumerate}
\item New York area: New York, Philadelphia, Baltimore, Washington;
\item Jacksonville area: Jacksonville, FL; 
\item Chicago  area: Chicago, Detroit, Indianapolis, Columbus, Columbus, Louisville; 
\item Dallas  area: Dallas, Fort Worth; 
\item Houston area: Houston, San Antonio, Austin; 
\item Denver area: Denver, 
\item Albuquerque area: Albuquerque, Santa Fe;
\item Phoenix area: Phoenix, El Passo, Tucson;
\item Los Angeles area: Los Angeles, San Diego; 
\item San Jose area: San Jose, San Francisco. 
\end{enumerate}

The attractor areas selected correspond to typical destinations of Mexican-born migrants (Fig.~\ref{methods}a) and also reflect current distribution of Mexican population in USA (Fig.~\ref{methods}b).  In each experiment we inoculated plasmodium between Ciudad Ju\'{a}rez and Nuevo Laredo. We decided to use a single inoculation site to avoid interference or even competition between plasmodium growing from several sites of inoculation.  The position of inoculation site chosen in experiments allows to cover most possible routes of migration, as illustrated in Fig.~\ref{methods}c.  

Containers with experimental setups were kept in closed yet naturally ventilated containers with water added to keep high humidity. We did not wetted the 3D terrain or covered it  with agar gel: the plasmodium grown on a bare Nylon surface.  The terrains were kept in darkness, at temperature 22-27$^\text{o}$C, except for observation and image recording. Configurations of plasmodium networks were photographed with FujiFilm FinePix 6000 camera and scanned with an Epson Perfection 4490 scanner.

\section{Results}

\begin{figure}[!tbp]
\centering
\subfigure[]{\includegraphics[width=0.4\textwidth]{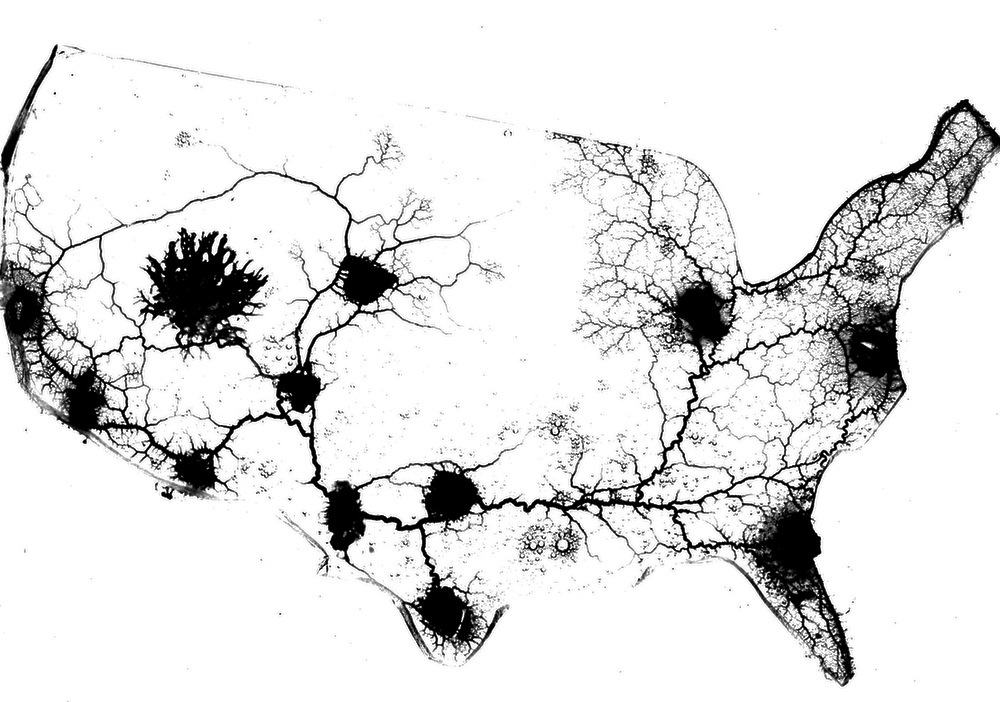}}
\subfigure[]{\includegraphics[width=0.4\textwidth]{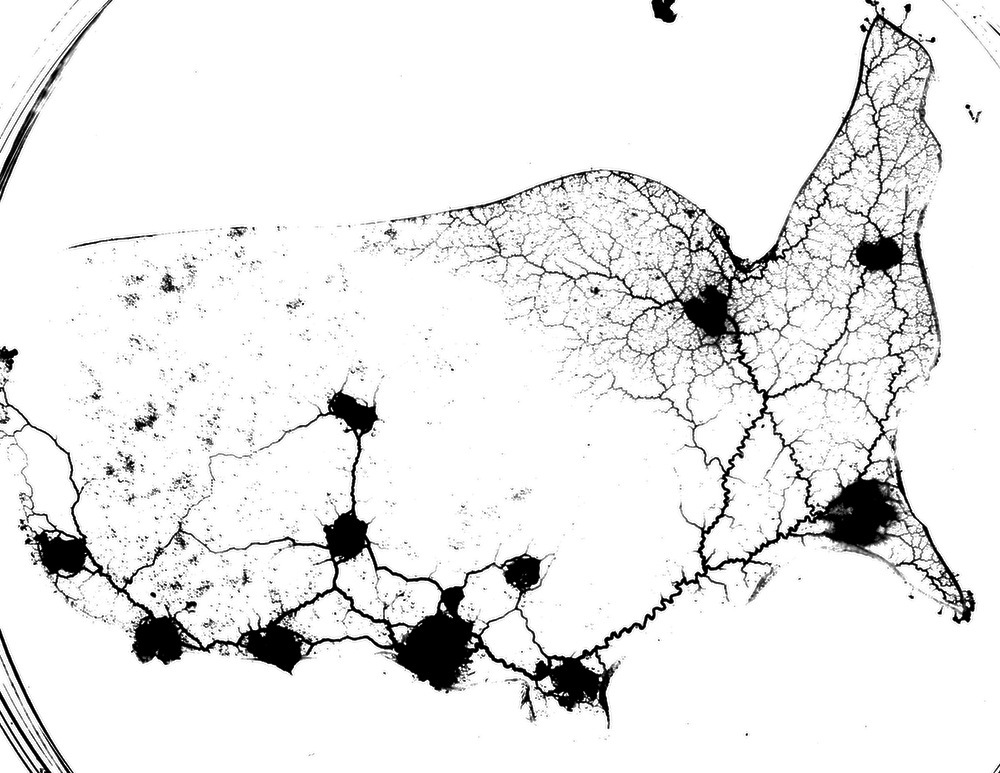}}
\subfigure[]{\includegraphics[width=0.4\textwidth]{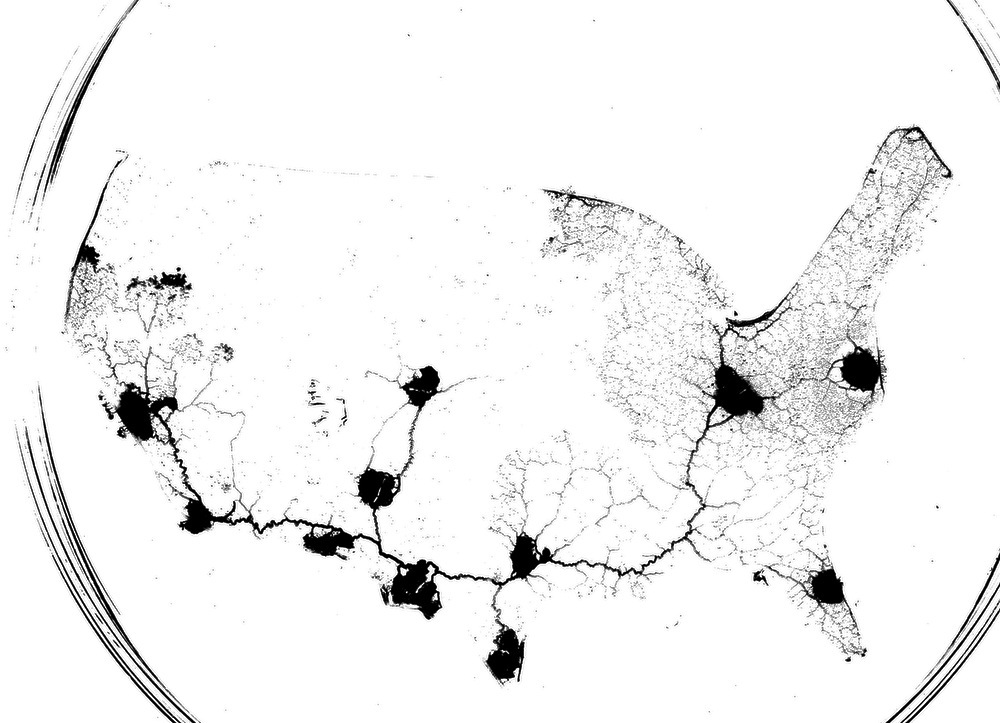}}
\subfigure[]{\includegraphics[width=0.4\textwidth]{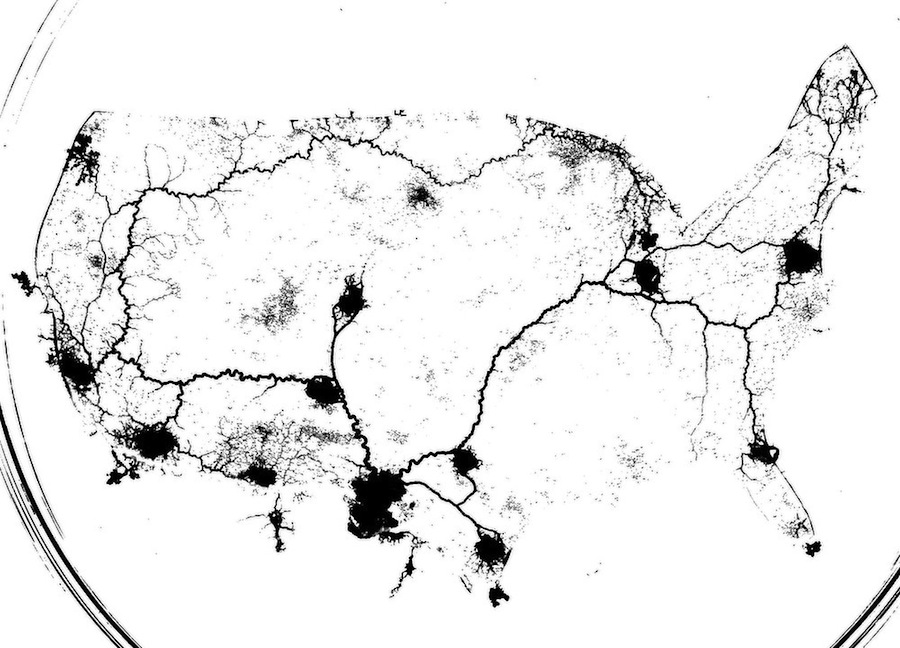}}
\subfigure[]{\includegraphics[width=0.7\textwidth]{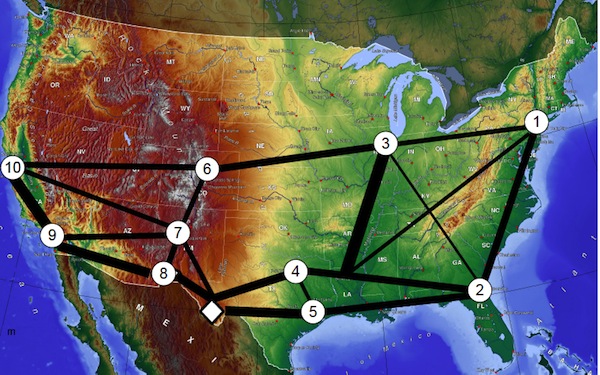}}
\caption{(a--d)~Snapshots of protoplasmic tubes spanning nutrients in $\mathbf{U}$. (e)~Weighted graph representing occurrences of slime mould links between areas of $\mathbf{U}$. Thickness of an edge is proportional to a number of trials where the edge is represented by a protoplasmic tube. Site of plasmodium inoculation is shown by rhomb.}
\label{flatagar}
\end{figure}

When colonising a flat substrate plasmodium is guided only by a configuration of attractant fields, generated by sources of nutrients, and possibly humidity distribution in the USA-shaped agar plates (Fig.~\ref{flatagar}). A typical scenario of colonisation is following. Plasmodium propagates from the inoculation site towards \Eight and either \Four or \Five. After colonising \Eight plasmodium grows towards \Seven and \Nine.  \Ten is usually colonised by slime mould which propagated from \Six or \Nine, and in few trials from \Seven (Fig.~\ref{flatagar}a--d). This matches well historical 
migration of Mexicans to Dallas or Houston areas, which are inhabited now by millions of Mexican migrants. Phoenix is 
a strategic point of migration to Los Angeles and Albuquerque areas.

When propagating east plasmodium grows from \Four or  \Five to \Two and then from \Two to \One. Also, very characteristically, the plasmodium makes a branching while propagating from \Four to \Two. Typically a branching site is located south of Louisiana. East branch continues towards \Two while north branch growth towards \Three. In few experiments we observed grows from the branching site to \One (Fig.~\ref{flatagar}).  A protoplasmic link between \One  and \Three is established either by plasmodium growing from \One to \Three or from \Three to \One. A transport route connecting \Three and \Six is usually developed by slime at the final stages of colonisation. Frequencies of routes observed in experimental laboratory trials are visualised in Fig.~\ref{flatagar}.  Chicago city is one of the most important cities for Mexican migrants and bears  substantial economically influence on Mexican community in USA. 
Migratory links between Chicago area and New York area serves social and economical needs of Mexican migrants for years; an anecdotal demonstration of the Chicago area's importance is an abundance of Spanish languages signs in 
transport and TV.

\begin{figure}[!tbp]
\centering
\subfigure[]{\includegraphics[width=0.45\textwidth]{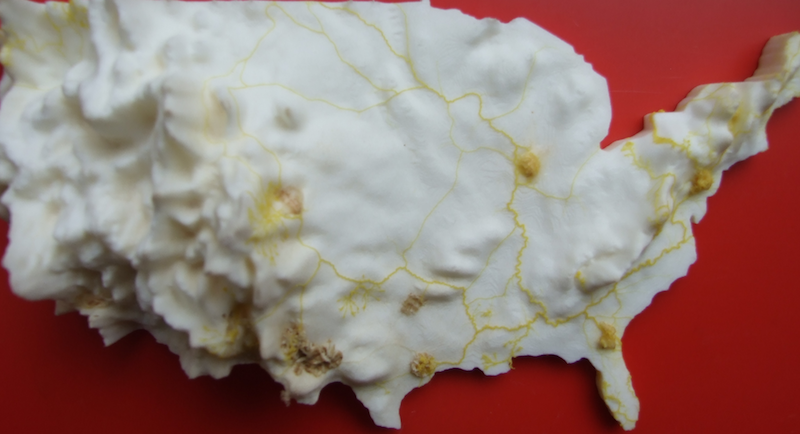}}
\subfigure[]{\includegraphics[width=0.45\textwidth]{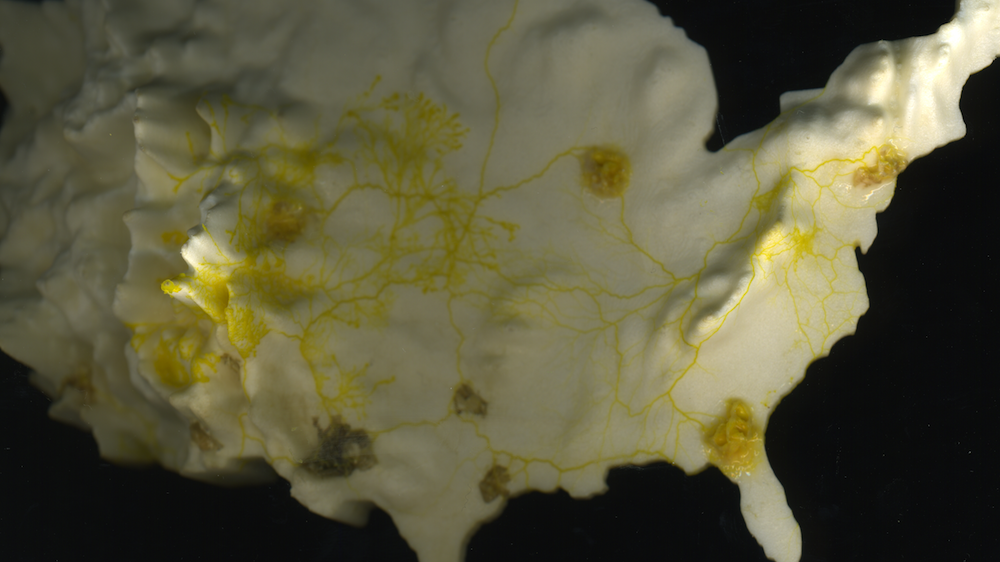}}
\subfigure[]{\includegraphics[width=0.95\textwidth]{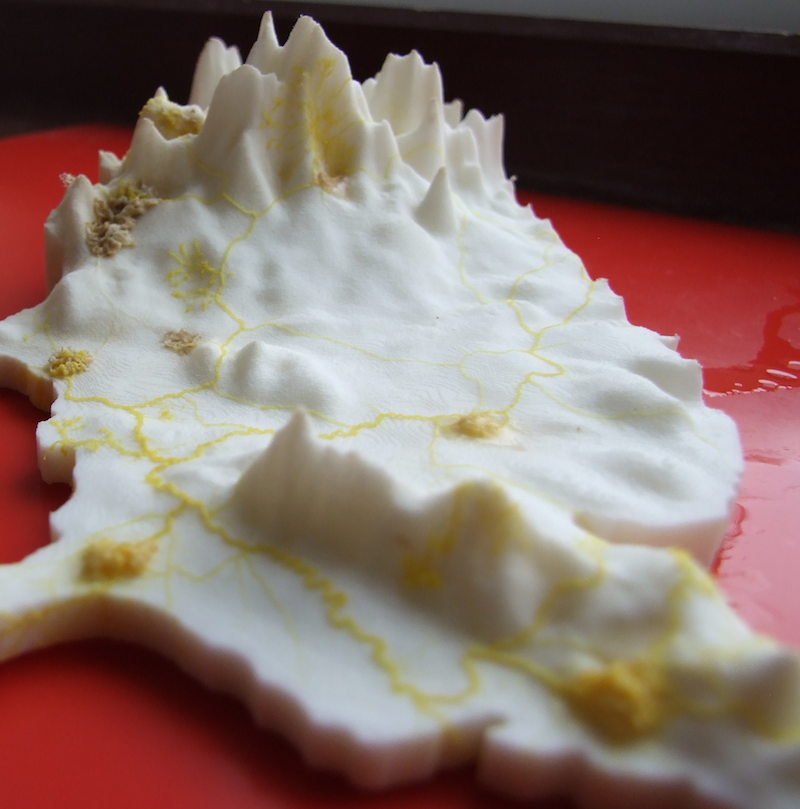}}
\caption{(a--c)~Photographs of 3D terrain of USA colonised by slime mould. }
\label{terrains}
\end{figure}

\begin{figure}[!tbp]
\centering
\includegraphics[width=0.8\textwidth]{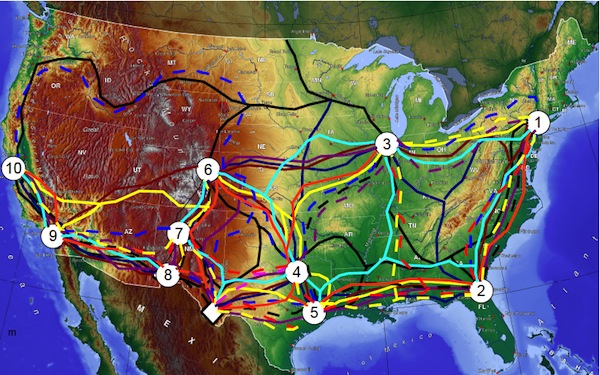}
\caption{Protoplasmic links developed in experiments with 3D terrain of USA. Links developed in each experiment are shown by a unique combination of colour and type of line. Site of plasmodium inoculation is shown by rhomb.}
\label{3Dscheme}
\end{figure}

Being placed on a 3D terrain with a source of nutrients slime mould would propagate towards the source of nutrients and navigate around elevations due to positive geotropism and relatively lower humidity of the elevations (Fig.~\ref{terrains}). The slime mould usually grows around mountains or cross mountains in the site of lowest elevation. Trajectories of protoplasmic tubes recorded in laboratory experiments are shown in Fig.~\ref{3Dscheme}. We can select three main stages of USA colonisation recorded in experiments with 3D terrain. Each stage takes 1-3 days of plasmodium growth.  At the first stage, plasmodium develops protoplasmic links from inoculation site to \Four, \Five, \Seven, and \Eight. At the second stage, slime mould propagates from \Four to \Five, or sometimes from  \Five to \Four; from \Five to \Two and \Three; from \Four to \Three and \Six;  from \Seven to \Six; and, from \Eight to \Nine.  At the third stage, the plasmodium grows from \Two to \One; from \Three to \Six and \One; from \Six to \Nine; and, from \Nine to \Ten. When all source of nutrients colonised the plasmodium may also develop a long-distant link between \Three and \Ten (Fig.~\ref{3Dscheme}).

\begin{figure}[!tbp]
\centering
\subfigure[]{\includegraphics[width=0.49\textwidth]{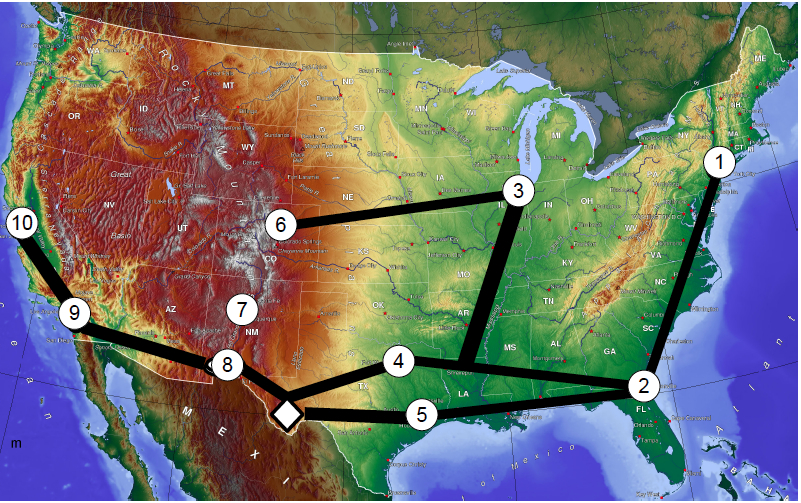}}
\subfigure[]{\includegraphics[width=0.49\textwidth]{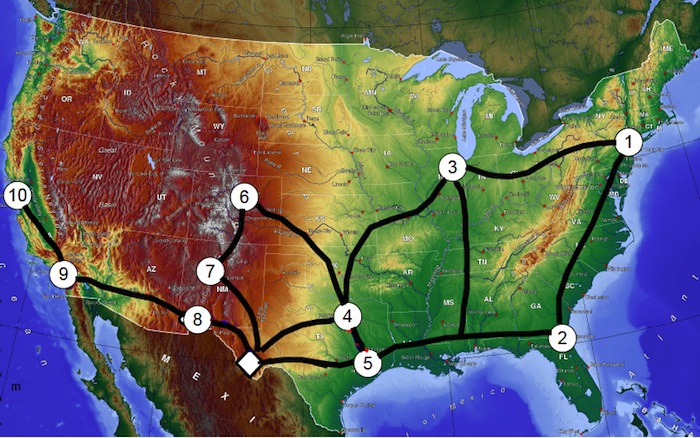}}
\subfigure[]{\includegraphics[width=0.49\textwidth]{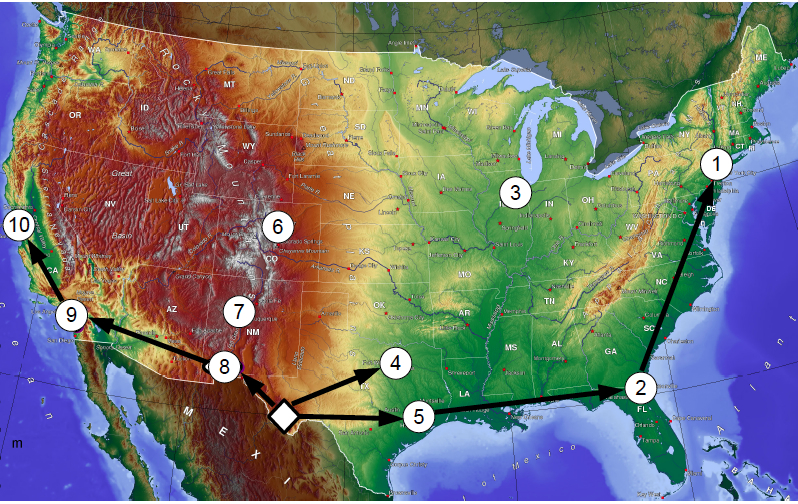}}
\caption{Strong components of migration graphs. (a)~at least in 6 out of 14 experiments (43\%) on flat agar. (b)~5 out of 12 experiments (42\%) on 3D terrain. (c)~a skeleton of migration: intersection of graphs (a) and (b), i.e. only edges represented in over 40\% of experiments on flat agar and 3D terrain are shown.}
\label{skeleton}
\end{figure}

Strong components, observed in over 40\% of laboratory experiments with flat agar and 3D terrain are shown in Fig.~\ref{skeleton}ab. In experiments with flat agar there are strong pathways from the entry point (inoculation site) to \Eight, \Five and \Four (Fig.~\ref{skeleton}a). From \Eight the migratory route develops to \Nine and then further north to \Ten. Migration from \Four and \Five proceed to \Two, and from \Two to \One. Also, there are well-established migratory pathways from \Four to  \Three (via branching point in south Louisiana), and from \Three to \Six  (Fig.~\ref{skeleton}a). 

Migratory routes developed by \emph{P. polycephalum} on a 3D terrain include most of the strong routes observed in experiments with flat substrate. Topology of protoplasmic routes reflect configurations of elevations in the routes' vicinity (Fig.~\ref{skeleton}b). Thus we witness remarkable avoidance of the Appalachian Mountains;  Sierra Nevada, Cascade Range, Rocky Mountains  and Ozark Plateau (Figs.~\ref{3Dscheme} and~\ref{skeleton}b). Also, additional strong pathways observed in experiments with 3D terrain include  links connecting \One and \Three, \Four and \Six, \Six and \Seven. 

A skeleton of migration links, with proposed directions of inward migration, is shown in Fig.~\ref{skeleton}c. The skeleton includes only edges observed in strong components of migratory graphs observed in both flat agar and 3D terrains experiments. The skeleton of migration consists of three components: links connecting the entry site to \Four, \Five and \Eight; west coast migration from \Eight to \Nine to \Ten; east cost migration from \Five to \Two and further to \One (Fig.~\ref{skeleton}c).

\section{Discussions}

Slime mould \emph{P. polycephalum} is now well recognised amorphous biological computer which works remarkable well in  experimental laboratory conditions. We presented an experimental laboratory prototype for analog modelling of human migration at a large scale. The prototype developed is a bionic devices, consisting of a living slime and a 3D plastic model of the USA. Experiments detailed in the paper can be classified as bio-computing,  future and emergent computing paradigms, bionics and analog computing and modelling. A protoplasmic network of slime mould \emph{Physarum polycephalum} transport nutrients and metabolites. The network  plays a role of a primitive distributed sensorial and nervous system, responsible for a choice of appropriate habitat and avoidance of harsh environmental conditions. Contents of the protoplasmic network is always in the flux. We speculate that the protoplasmic network is a good biological model of a migration network emerged in a transnational communities of people living between Mexico and USA~\cite{conway_cohen_1998}. 

In laboratory experiment with the slime mould we imitated Mexican migration on a flat agar and 3D Nylon terrain. Migratory links developed on a flat substrate may represent air transportation while routes on 3D terrain signify ground transportation. We found that the plasmodium imitates human movements amongst elevations because \emph{P. polycephalum} is  gravisensitive and positively geotropic~\cite{block_1986,block_1998} and geopolar~\cite{block_1989}.

The experimental setup adopted well represents principle mechanics of migration~\cite{massey_espinosa_1997}:
\begin{itemize}
\item  To imitate a social capital formation we place oat flakes in areas with high population of Mexican migrants. 
\item  Human capital formation is imitated by \emph{P. polycephalum} via self-reinforcing of protoplasmic tubes: ticker tubes become thicker and thinner become weaker. 
\item Market consolidation trends are `hard-wired' in the slime mould behaviour via its foraging activity. 
\end{itemize}
Strong component of migration routes extracted from laboratory experiments, represent all types of migration~\cite{pries_2001}: 
\begin{itemize}
\item `internal' migration to North Mexican ex-territories (routes \Eight to \Nine and \Ten); 
\item diaspora migrations, e.g. links between \Six and \Seven, \One and \Three,  \Four and \Five, \One and \Two, \Three and \Four,  \Two and \Five; 
\item returned migration, e.g. links connecting the migrants' entry point to \Eight,
\Seven, \Four and \Five. 
\end{itemize}

\section{Acknowledgement}

This work was supported by the EU research project ``Physarum Chip: Growing Computers from Slime Mould" (FP7 ICT Ref 316366).

\end{document}